\documentclass[prl,twocolumn]{revtex4}
\newcommand{\ket}[1]{|#1\rangle}
\newcommand{\bra}[1]{\langle#1|}

\usepackage{graphicx}

\begin{document}

\title{Experimental violation of a spin-1 Bell inequality using maximally-entangled four-photon states}

\author{John C. Howell, Antia Lamas-Linares and D. Bouwmeester}

\address{
Centre for Quantum Computation, Clarendon Laboratory, University
of Oxford,\\ Parks Road, OX1 3PU Oxford, United Kingdom}

\begin{abstract}
We demonstrate the first experimental violation of a spin-1 Bell
inequality.  The spin-1 inequality is a calculation based on the
Clauser, Horne, Shimony and Holt formalism.  For entangled spin-1
particles the maximum quantum mechanical prediction is 2.552 as
opposed to a maximum of 2, predicted using local hidden variables.
We obtained an experimental value of 2.27 $\pm 0.02$ using the
four-photon state generated by pulsed, type-II, stimulated
parametric down-conversion. This is a violation of the spin-1 Bell
inequality by more than 13 standard deviations.
\end{abstract}
 \maketitle

\begin{figure}
\scalebox{0.5}{\includegraphics{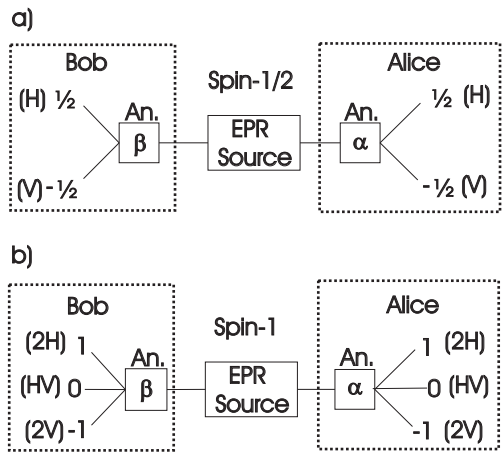}} \caption{(a) A
Bohm-type spin-1/2 Gedanken experiment. Alice and Bob each receive
one particle from an entangled pair. Both Alice and Bob measure
either $+1/2$ or $-1/2$ for any orientation of their analyzers
($\alpha$ and $\beta$) and their outcomes will be correlated
according to quantum mechanical predictions. (b) Bohm-type spin-1
Gedanken experiment. Similarly Alice and Bob measure correlated
values of $1$, $0$ and $-1$.} \label{SpinAnalyzers}
\end{figure}

In 1935 Einstein, Podolsky and Rosen (EPR) showed that quantum
mechanics implied nonlocality \cite{Einstein35}.  In 1951, Bohm
discussed correlations of two entangled spin-1/2 particles of the
form $\frac{1}{\sqrt{2}} \left( \vert + \frac{1}{2},-\frac{1}{2}
\rangle- \vert -\frac{1}{2} ,+\frac{1}{2}\rangle \right)$.  A
graphical representation of entangled spin-1/2 Bohm-type apparatus
is shown in Fig. \ref{SpinAnalyzers} a).  One spin-1/2 particle is
sent to Alice who analyzes the particle in a basis determined by
her analyzer orientation $\alpha$.  The conjugate particle is sent
to Bob who analyzes his particle in a basis determined by his
analyzer orientation $\beta$.  Alice and Bob then measure either
$+1/2$ or $-1/2$.  Quantum mechanics predicts that Alice's and
Bob's measurements are correlated such that they appear to violate
EPR's notion of locality.  In 1965, Bell \cite{Bell65} and later
Clauser, Horne, Shimony and Holt (CHSH) \cite{CHSH} used the
spin-1/2 Bohm-type ideas to show that quantum predictions and
local explanations of the correlations were mathematically
incompatible. Since that time spin-$1/2$ experiments employing
polarization entangled photons
\cite{Clauser78,Aspect81,Kwiat95,Weihs98}, momentum entangled
photons \cite{Kwiat93,Brendel98} and most recently with trapped
ions \cite{Rowe01} have been used to verify the quantum
predictions. The nonlocality of these entangled spin-1/2 particles
has been employed in several important applications in the field
of quantum information such as dense coding \cite{Mattle96},
quantum cryptography \cite{Ekert91,Naik00} and quantum
teleportation \cite{Bennett93}.

A natural extension of the research on entangled particles is the
study of entangled states of spin-$s$ objects ($s>1/2$). Gisin and
Peres showed that entangled particles with arbitrary large spins
still violated a Bell inequality\cite{Gisin92}. This result
implies that large quantum numbers are no guarantee of classical
behavior. Apart from its the fundamental interest
\cite{Gisin92,Cirel'son80,Kaszlikowski00}, entangled states of
spin-$s$ objects are of clear interest for applications in quantum
information due to the higher dimensional Hilbert space associated
to these states (e.g. quantum cryptography, dense coding and bound
entanglement\cite{horodecki98}). Despite the strong interest
expressed in entangled spin-$s$ objects no experimental
realization has been reported to date. Here we present the first
experimental demonstration of a violation of the Bell inequality
for entangled spin-1 objects. We use the fact that the
polarization entangled four-photon fields (2-photons in each of
two spatial modes) of pulsed parametric down-conversion are
formally equivalent to two maximally entangled spin-1 particles
\cite{lamas01}. This is related to theoretical work by Drummond
\cite{drummond83} in which he describes cooperative emission of
wavepackets containing $N$-bosons and proved that multiparticle
states could violate the Bell inequalities. The connection between
states produced in parametric down-conversion and the $N$-boson
multiparticle states has recently been discussed by Reid \emph{et
al} \cite{reid01}.

A spin-1 particle would have three distinct measurement states
($\vert1\rangle$,$\vert0\rangle$, or $\vert-1\rangle$) The spin-1
analog of Bohm's entangled spin-1/2 particles would be given by
\begin{equation}
|\Psi_1\rangle=\frac{1}{\sqrt{3}}\left(|1,-1\rangle-|0,0\rangle+|-1,1\rangle
\right) \label{Spin1}
\end{equation}
A graphical representation of Bohm-type entangled spin-1 particles
is shown in Fig. \ref{SpinAnalyzers} b).  Similar to the spin-1/2
case, one spin-1 particle is sent to Alice who analyzes the
particle in a basis determined by her analyzer orientation
$\alpha$.  The conjugate particle is sent to Bob who analyzes his
particle in a basis determined by his analyzer orientation
$\beta$.

The crux of Bell inequalities is that the probabilities $P$ of the
``locally explicable" outcomes can be decoupled as
\begin{equation}
P(A,B\vert \alpha,\beta)=P(A\vert \alpha,\lambda)P(B\vert
\beta,\lambda)
\end{equation}\label{crux}
where $\lambda$ accounts for all local hidden variables. $A$ and
$B$ refer to the measurement results
($\vert1\rangle$,$\vert0\rangle$, or$\vert-1\rangle$) obtained by
Alice and Bob using analyzer orientations $\alpha$ and $\beta$
respectively. Taking a local realists point of view we assume that
the measurement outcomes can be decoupled. We define a spin-1
local hidden variable measurement combination
\begin{equation}
E^{HV}(\alpha,\beta)= \int d\lambda f(\lambda)
\overline{A}(\alpha,\lambda)\overline{B}(\beta,\lambda).
\end{equation}
where
\begin{eqnarray}
\overline{A} (\alpha,\lambda) = P(1\vert\alpha,\lambda) - P(0\vert
\alpha,\lambda) + P(-1\vert \alpha,\lambda)
\\
\overline{B} (\beta,\lambda)=P(1\vert \beta,\lambda)-P(0\vert
\beta,\lambda)+P(-1\vert \beta,\lambda)
\end{eqnarray}
Because the signs of the probabilities in both $\overline{A}
(\alpha,\lambda)$ and $\overline{B} (\beta,\lambda)$ are
different, it must be true that $\vert \overline{A}
(\alpha,\lambda) \vert \leq 1$ and $\vert \overline{B}
(\beta,\lambda) \vert \leq 1$. The derivation of the spin-$1$ Bell
inequality proceeds exactly as the spin-$1/2$ formalism
\cite{CHSH,Bellbook}, leading to
\begin{equation}
S=\vert
E(\alpha,\beta)-E(\alpha,\beta')+E(\alpha',\beta)+E(\alpha',\beta')\vert
\leq 2 \label{BellInequality}
\end{equation}
Hence, the maximum possible value that can be achieved, assuming
locally explicable outcomes is 2. On the other hand, quantum
mechanics states that the measurement probabilities on Alice's and
Bob's side cannot be decoupled, which implies that the quantum
mechanical measurement outcome is
\begin{eqnarray}
E^{QM}(\alpha,\beta)&=&P(1,1\vert \alpha,\beta)- P(1,0\vert
\alpha,\beta)\nonumber\\&+&P(1,-1\vert \alpha,\beta)-P(0,1\vert \alpha,\beta) +P(0,0\vert \alpha,\beta)\nonumber \\
&-& P(0,-1\vert \alpha,\beta)+P(-1,1\vert
\alpha,\beta)\nonumber\\&-&P(-1,0\vert \alpha,\beta)+P(-1,-1\vert
\alpha,\beta) \label{QuantumPrediction}
\end{eqnarray}
Using the Bell inequality in eqn. (\ref{BellInequality}) we obtain
a theoretical maximum violation of 2.552 in agreement with Gisin
and Peres \cite{Gisin92}. This prediction was obtained using
analyzer rotations of $\alpha=0^o$, $\alpha'=22.5^o$,
$\beta=11.25^o$, and $\beta'=33.75^o$.

The entangled quanta we wish to use are the multi-photon modes of
a polarization entangled field \cite{lamas01} produced by pulsed
type-II parametric down-conversion. The first order term of
parametric down-conversion is $1/\sqrt{2}\left(\vert H,V\rangle-
\vert V,H\rangle\right)$, which is used in spin-$1/2$ Bell
inequality experiments. A graphical representation is shown in
parentheses in Fig. \ref{SpinAnalyzers} a). However we are
interested in the second order term of the down-converted field.
By using postselection we can selectively measure this term, which
is given by
\begin{equation}
\frac{1}{\sqrt{3}}\left(\vert 2H,2V\rangle -\vert HV,VH\rangle
+\vert 2V,2H\rangle \right)
\label{maximal4}
\end{equation}
where the first term in the kets represent the polarization of the
photons sent to Alice and the second term in the kets represent
the polarization of the photons sent to Bob. For example, the
$\vert 2H,2V\rangle$ means that if Alice measures two horizontal
photons, then Bob will measure two vertical photons. The photons
sent to Alice have three possible measurement outcomes with equal
probabilities, namely $\vert 2H\rangle$, $\vert HV\rangle$ and
$\vert 2V\rangle$, which we will define as the $\vert 1\rangle$,
$\vert 0\rangle$ and $\vert -1\rangle$ state respectively (see
fig. \ref{SpinAnalyzers}).  Thus, it is \textit{not} the photons
that are the spin-1 particles, but the two-photon polarization
entangled modes.

\begin{figure}
\scalebox{0.4}{\includegraphics{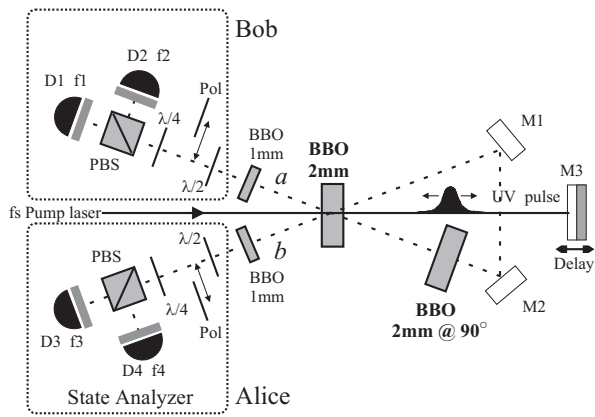}}
\caption{Experimental setup for generation and detection of
entangled spin-1 singlets. A type-II non-collinear parametric
down-conversion process creates four photon states which are
amplified by the double pass configuration. The detection is done
at Alice's and Bob's sides by postselection as described in the
main text.} \label{Setup}
\end{figure}

A schematic of our experimental setup is shown in Fig.
\ref{Setup}. The pump laser is a 120 fs pulsed, frequency doubled,
Ti:Sapphire laser operating at 390 nm with an 80 MHz repetition
rate.  The pump enters a nonlinear beta-barium borate (BBO)
crystal cut for type-II phase matching \cite{Kwiat95}. The
down-converted field is then fed back into the crystal along with
the retro-reflected pump beam. The difference in the round-trip
path length of the pump beam and down-converted field is much
smaller than the coherence length of the 5nm bandwidth frequency
filtered down-converted photons. The feedback loop for the
entangled fields contains a 2 mm BBO crystal rotated 90$^o$ with
respect to the optical axis of the down-conversion crystal, which
compensates for the temporal walk off. Such alignment yields very
good spatial and temporal overlap with which-pass interference
visibilities of $98\%$.

The primary purpose for using the two-pass scheme is to increase
the count rates.  For pulsed four-photon down conversion the count
rates increased by a factor of 16 for two passes as opposed to 1
pass, provided that both down-conversion fields are exactly in
phase and completely indistinguishable. This leads to
approximately 10 four-photon coincidence detections per second. To
perform active stabilization of the phase we use the fact that
under the same conditions there is maximum constructive
interference for the much more intense two-photon state (singlet
spin-$\frac{1}{2}$). Thus the two-photon coincidences can then act
as a precision, low-noise four-photon intensity reference.

The analysis setup is shown in the dashed box in Fig. \ref{Setup}.
Each analyzer contains a $\lambda/2$ waveplate, a polarizer, a
$\lambda/4$ waveplate, a polarizing beam splitter (PBS), narrow
bandwidth filters (5 nm) and 2 single photon detectors. The
half-wave plates are used to set the desired $\alpha$ and $\beta$
on Alice's and Bob's sides. The $\vert HV\rangle$ state is
detected by having the quarter wave plate oriented $0^o$ with
respect to the horizontal polarization axis. The photons then pass
through the quarter wave plate unaltered and are split up at the
PBS. The $\vert 2H\rangle$ state is measured by inserting a linear
polarizer oriented such that only horizontally polarized photons
are transmitted. The quarter wave plate rotated by $45^o$ with the
PBS is an effective 50/50 beam splitter. Thus, the probability for
measuring two photons (one in each detector) on Alice's or Bob's
side is reduced by a factor of 2 due to the binomial measurement
statistics. In addition, inserting a polarizer introduces
additional unavoidable losses in the mode and further reduces the
probability of measurement compared to that of the $\vert HV
\rangle$ state. It was experimentally determined that the
two-photon measurement probability of the $\vert 2H\rangle$ state
was 43.1 \% on Alice's side and 43.4 \% on Bob's side compared to
50\% for an ideal 50/50 beam splitter and lossless polarizer.
Measuring the $\vert 2V\rangle$ is the same as the $\vert
2H\rangle$ except that the polarizer is rotated by 90$^o$.

With the configuration just described, it is necessary to measure
36 probabilities, nine from eqn. (\ref{QuantumPrediction}), for
each of the four analyzer settings in eqn. (\ref{BellInequality}).
The experimental results for one analyzer setting (namely
$\alpha=-16^o$,$\beta'=14^o$) are listed in the table below. The
measurements were taken by observing the raw 4-fold coincidence
counts of all nine measurement possibilities. Each data point is
the average over twelve 60 second intervals. The data obtained
using two polarizers were then multiplied by a factor
$1/(0.431)(0.434)$. The data obtained using a polarizer on Alice's
(Bob's) side were multiplied by a factor of $1/(0.431)$
($1/(0.434)$). This modified data is shown under the Mod. column
and the corresponding probability under Prob. Similar tables have
been measured for the other three analyzer orientations (using
$\alpha'=4^o$, $\beta=6^o$). Combining all this data we arrive at
a single value, $S=2.27\pm0.02$.

\vspace{0.5cm}

\begin{tabular}{|c|c|c|c|}

  \hline
   $\alpha=-16^o$, $\beta'=14^o$ & $\langle Counts (60s)\rangle$ & Mod. & Prob. \\ \hline
  P(1,1) &  2.20& 11.71 & 2.25\% \\
  P(1,-1) & 18.04 & 96.05 & 18.46\% \\
  P(-1,1) &  17.37& 92.48 & 17.77\% \\
  P(-1,-1) & 1.78 & 9.48 & 1.82\% \\
  P(1,0) & 21.92 & 50.47 & 9.70\% \\
  P(0,1) & 33.67 & 77.86 & 14.96\% \\
  P(-1,0) & 21.43 & 49.34 & 9.48\% \\
  P(0,-1) & 28.74 & 66.46 & 12.77\% \\
  P(0,0) & 66.50 & 66.50 & 12.78\% \\ \hline
  Total & - & 520.35 & 100\% \\ \hline
  \end{tabular}

\vspace{0.5cm}

\begin{figure}
\rotatebox{-90}{\scalebox{0.35}{\includegraphics{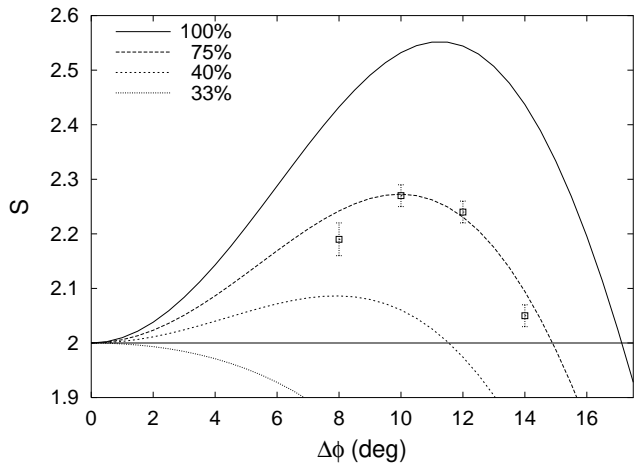}}}
\caption{The value of $S$ is plotted as a function of the angle
difference between analyzer axes. The curves correspond to
different entanglement visibilities. Our experiment had $75\%$
entanglement visibility and the experimental points are shown
along with the corresponding theoretical prediction.}
\label{fig:QuantumPrediction}
\end{figure}

Due to the rotational symmetry of the maximally entangled state
described in eqn. (\ref{maximal4}), the only relevant experimental
setting to obtain a maximum violation, is the difference in angles
between the analyzer settings
($\Delta\phi=\beta-\alpha=\alpha'-\beta=\beta'-\alpha'$). However,
we do expect some degree of mixture in the state created in our
source. A simple model for the noise in our source is given by a
statistical mixture of the individual terms of the pure state. The
resulting density matrix is given by
\begin{eqnarray}
\rho&=&p\left(\ket{\psi_{pure}}\bra{\psi_{pure}}\right)+\frac{(1-p)}{3}(
\ket{2H,2V}\bra{2H,2V}\nonumber\\&+&
\ket{HV,VH}\bra{HV,VH}+\ket{2V,2H}\bra{2V,2H})
\end{eqnarray}
where $p$ is the probability of having the pure entangled state.
This model is a simple function of only one parameter ($p$), which
is directly related to the lowest fringe visibility obtained by
fixing the polarization orientation of one analyzer and rotating
the other. This is often referred to as entanglement visibility.
The presence of noise will degrade this entanglement visibility
and break the rotational symmetry. Because of this symmetry
breaking, it will no longer be irrelevant where we define the zero
of the analyzers. In particular we want to set our measurement
axes ($\alpha,\beta,\alpha',\beta'$) such that they are symmetric
around the origin defined by the optical axis of the
down-conversion crystal; in this way we minimize the effect of the
classical noise in achieving a Bell violation. The maximum
violation for a given level of noise occurs at a reduced angle
difference $\Delta\phi$ compared with the ideal noiseless case
\cite{eberhard}. The curves in fig. \ref{fig:QuantumPrediction}
are calculated values of $S$ as a function of the angle difference
$\Delta\phi$ for various levels of noise corresponding to the
indicated entanglement visibilities. For a 100\% visibility the
maximum value of $S$ is 2.55. In our experiment we measured an
entanglement visibility of $75\%$ which --in our modelling of the
noise-- corresponds to $p=0.69$ and has a maximum of $S=2.28$ for
$\Delta\phi=10$. This is in good agreement with our measured value
of $S=2.27\pm0.02$ at $\Delta\phi=10^o$. This is more than $13$
standard deviations away from the maximum value explainable by
local realistic theories, $S=2$. In order to rule out systematic
errors we measured three additional points along the curve of 75\%
visibility. Each of these also violates the Bell inequality as
expected.

Since our entangled spin-1 objects are constructed from four
photons produced by parametric down-conversion, it is natural to
consider whether the observed violation could be explained in
terms of a product state of spin-1/2 pairs. However we have
performed calculations showing that the four-photon state created
in this way does not violate the inequality in
\ref{BellInequality}.

Stimulated emission, which has been used to enhance our
four-photon counts, also enhances Bell-type experiments by
improving mode matching of the source to the detectors.  We
observed an increase in detection efficiency from 12\% for a
single pass to 18\% for a double pass. With many passes through
the crystal \cite{Lamas} and improved coupling/detection
optics\cite{Kurtsiefer}, it might be possible to obtain
efficiencies high enough to perform completely loophole-free
violations of Bell inequalities.

In summary, we have reported the first experimental violation of a
spin-1 Bell inequality. The experimentally determined value was
2.27 $\pm$ 0.02 which is in excellent agreement with the value of
2.28 expected from the entanglement visibility of 75\%. In
principle, the method can be extended to higher spin numbers.
These results open up the exploration of spin-1 (and higher)
states for optical quantum information.

\noindent {\bf Acknowledgements}

We would like to thank A. Ekert and A.V. Sergienko for helpful
discussions. This work was supported by the EPSRC GR/M88976 and
the European QuComm (ISI-1999-10033) projects.

\end{document}